\documentclass[pra,aps,12pt,tightenlines,showkeys,amsmath,amssymb]{revtex4-1}

\usepackage[english]{babel}
\usepackage{afterpage}
\usepackage{graphicx}
\usepackage{dcolumn}
\usepackage{bm}
\usepackage{color}
\usepackage{slashed}
\usepackage{latexsym}

\newcommand\numberthis{\addtocounter{equation}{1}\tag{\theequation}}

\usepackage{anysize}
\usepackage{float}
\usepackage{perpage}
\usepackage{dsfont}
\usepackage{xcolor}
\usepackage[12pt]{moresize}
\usepackage[title]{appendix}

\newcommand{\kolorb}[1]{{\color[rgb]{0,0,0.8} #1}}

\begin{document}

\title{Vortex Structures and Momentum Sharing\\ in Dynamic Sauter--Schwinger Process}
\author{A. Bechler}
\email[E-mail address:\;]{Adam.Bechler@fuw.edu.pl}
\author{F. Cajiao V\'elez}
\author{K. Krajewska}
\author{J. Z. Kami\'nski}

\affiliation{Institute of Theoretical Physics, Faculty of Physics, University of Warsaw, Pasteura 5,
02-093 Warszawa, Poland 
}

\date{\today}

\begin{abstract}
Vortex pattern formation in electron-positron pair creation from vacuum by a time-dependent electric field of linear polarization is analyzed.
It is demonstrated that in such scenario the momentum distributions of created particles
exhibit vortex-antivortex pairs. Their sensitivity to the laser field parameters such as the field frequency and intensity is also
studied. Specifically, it is shown that with increasing field frequency accross the one-photon threshold additional vortex-antivortex 
pairs appear. Their location in the momentum space is consistent with a general threshold behavior of probability distributions of created
electrons (positrons). Namely, while for small field frequencies the particles tend to be created along the field polarization direction, for 
large enough frequencies they are predominantly generated in the perpendicular direction. Such change in longitudinal and transverse
momentum sharing of created particles occurs accross the one-photon threshold.

\end{abstract}



\maketitle

\newpage

\section{Introduction}
\label{sec::intro}

Nonlinear response of quantum vacuum to macroscopic electromagnetic fields, leading to creation of
electron-positron ($e^-e^+$) pairs, has been predicted by Sauter~\cite{Sauter}, Heisenberg and Euler~\cite{Heisenberg-Euler}, 
and Schwinger~\cite{Schwinger}. Since then, various authors have made significant contributions 
to our current understanding of this process, which we will refer to as the Sauter-Schwinger process. Specifically, Bia\l ynicki-Birula, G\'ornicki and Rafelski have established 
a new framework for treating the quantum vacuum in electromagnetic fields~\cite{IBB1} (see, also, Refs.~\cite{IBB102,IBB103} and the PhD thesis of {\L}. Rudnicki \cite{LR2012}). 
This is by means of what they called the Dirac-Heisenberg-Wigner (DHW) function, which describes the $e^-e^+$ densities in phase space. 
Later on, the method was 
largely explored for the case of spatially homogeneous electric fields (see, e.g.,~\cite{HAG,IBB2,archiwum,Blinne,Blinne1,PhDthesis,Xie1,Xie2,IZBB}). For instance, the quantum kinetic approach was recovered in
that case~\cite{HAG} and various analytical results for exactly solvable fields were derived~\cite{HAG,IBB2,archiwum}. Recently, the spontaneous formation
of time-crystal structures in the $e^-e^+$ pair creation was discovered by Bia\l ynicki-Birula and Bia\l ynicka-Birula in Ref.~\cite{IZBB}. Other applications of the DHW formalism in the context
of pair creation concern the case of parallel spatially homogeneous electric and magnetic fields~\cite{Sheng}, the standing electric 
wave~\cite{Heben,Berenyi,Naseri,Mohamedsedik}, and inhomogeneous electric and magnetic fields in one spatial direction~\cite{Alkofer1,Kohlfu1,Kohlfu2}. 
The latter limitation follows entirely from performing capabilities of current computers, as the DHW method is very general and can be used 
in arbitrary dimensions. It is also important to emphasize that the DHW method is not limited to describing pair creation from vacuum.
For instance, it was argued that DHW is very useful for practical plasma applications such as studies of
Langmuir waves in a high-density plasma~\cite{plasmareview}.

Another area of research where prof. I. Bia\l ynicki-Birula has largely contributed to is related to quantum vortices. It follows from the hydrodynamical
formulation of quantum mechanics that the probability fluid can inherently possess vortices~\cite{Cieplak}. They are defined as phase singularities of the 
wave function and their strenght is measured in terms of a topological charge~\cite{Cieplak,BS,Taylor}. As it was discussed in~\cite{BS,Taylor}, the vortices
form isolated lines, that either emerge from a single point forming a closed loop or can be created as a pair of lines with opposite topological
charges. Those mechanisms of creation and subsequent annihilation of vortex-antivortex pairs were confirmed recently in a series of papers focused on
vortex structures in strong-field ionization~\cite{Pe2,Pe1,Pe3,Felipe,Majczak}. Specifically, it was demonstrated that the vortex structures are very sensitive to
the laser field parameters, and so they can be easily steered by the field. While the aforementioned papers deal with quantum vortices in nonrelativistic 
quantum mechanics, their notion can be also extended to relativistic quantum theory as has been proposed by Bia\l ynicki-Birula and Bia\l ynicka-Birula
in~\cite{BBvortices}. See also, the construction of knotted vortex states, or hopfionslike states, in relativistic quantum 
mechanics~\cite{BBhopfions} by the same authors.

Note that electron-positron pair creation and ionization are formally similar, as they are both threshold-related phenomena which can be driven by 
external dynamically changing fields. For this reason, one might expect similar effects being exhibited in both processes. Keeping this in mind, in the 
current paper we investigate whether vortex structures similar to~\cite{Pe2,Pe1,Pe3,Felipe,Majczak} can be observed in the probability amplitude of $e^-e^+$ pair
creation in the presence of a linearly polarized time-dependent electric field. Our emphasis will be on threshold behavior of those patterns, 
which can be studied, for instance, by changing the frequency of the driving field. As we will show, this is in agreement with the longitudinal
and the transverse momentum sharing of created particles across the threshold which has been studied in Ref.~\cite{KK012104}.
At this point we would like to mention that other structures, known as
spiral vortex patterns, were found in strong-field ionization~\cite{SAF,Wollen} and later on in pair production~\cite{Xie1,Xie2} for certain combinations
of circularly polarized electric field pulses. As demonstrated however in Ref.~\cite{Pe3}, in the case of ionization such spirals in momentum
distributions of photoelectrons does not necessarily carry a nonzero topological charge which distinguishes them from vortices analyzed in 
Refs.~\cite{Pe2,Pe1,Pe3,Felipe,Majczak}. The same is expected to hold for pair creation.

Our paper is organized as follows. Based on the original derivation presented in Ref.~\cite{IBB1}, we introduce the DHW formalism in Sect.~\ref{theory}. 
Bispinorial decomposition of the DHW-function for a spatially homogeneous electric field is presented in Sect.~\ref{bispinorial} and 
the final equations for a linear polarization are given in Sect.~\ref{linear}. Section~\ref{sec:threshold} is devoted to vortex patterns 
in $e^-e^+$ pair creation and their sensitivity to external field parameters, especially when passing across the threshold. Another threshold-related
effect is discussed in Sect.~\ref{sec:lt}, where we demonstrate how the particles momentum is redistributed across the threshold of pair creation. 
Our final remarks are given in Sect.~\ref{sec:conclusions}.

\section{The DHW-function for fermion field}
\label{theory}
	The DHW-function for the fermion field is defined as~\cite{IBB1}
\begin{align}\label{0.1}
	W_{\alpha\beta}(\bm{x},\bm{p},t)=-\frac{1}{2}\int d^3se^{-i\bm{p}\cdot\bm{s}}\langle 0|\mathcal{U}(\bm{s},\bm{x},t)[\Psi_{\alpha}(\bm{x}+\bm{s}/2,t),\Psi^\dag_\beta(\bm{x}-\bm{s}/2,t)]|0\rangle,
\end{align}
where  the factor $\mathcal{U}(\bm{s},\bm{x},t)$ contains line integral of the vector potential in temporal gauge $A^0=0$,
\begin{align}\label{0.2}
	\mathcal{U}(\bm{s},\bm{x},t)=\exp\left[-ie\int_{-1/2}^{1/2}d\xi\bm{s}\cdot\bm{A}(\bm{x}+\xi\bm{s},t)\right],
\end{align}
and assures gauge invariance of the DHW-function, whereas $\Psi_\alpha,\,\Psi_\beta$ are the fermion field operators in Heisenberg picture. We use 
here the version of DHW-function with the vacuum expectation value~\cite{HAG}; in general, however any pure or mixed state can be used~\cite{IBB1}. 
The DHW-function is a $4\times 4$ Hermitian matrix and as such can be decomposed in terms of 16 Hermitian matrices $\Gamma_a$ with real coefficients 
depending, in general, on $\bm{x}$, $\bm{p}$ and time. Matrices $\Gamma_a,\,a=0,1,2,...,15$  can be constructed as Kronecker products of two sets of 
Pauli matrices (including the identity matrix), $(I_2,\rho_j)$ and $(I_2,\sigma_j)$~\cite{IBB1}. The correspondence looks as follows
\begin{align}\label{0.3}
	\begin{split}
		&	\Gamma_0=I_4,\,\Gamma_j=\rho_j\otimes I_2,\,\Gamma_{j+3}=I_2\otimes\sigma_j,\,\Gamma_{j+6}=\rho_1\otimes\sigma_j,\\ &\Gamma_{j+9}=\rho_2\otimes\sigma_j,\,\Gamma_{j+12}=\rho_3\otimes\sigma_j,
	\end{split}
\end{align}
where the index $j=1,2,3$. In terms of standard $\gamma$-matrices,
\begin{align}\label{0.4}
	\begin{split}
		& \Gamma_0=I_4,\,\Gamma_1=\gamma_5,\,\Gamma_2=-i\gamma^0\gamma_5,\,\Gamma_3=\gamma^0,\,\Gamma_{j+3}=\Sigma^j,\,\\
		&\Gamma_{j+6}=\alpha^j,\,\Gamma_{j+9}=-i\gamma^j,\,\Gamma_{j+12}=\gamma^0\Sigma^j, 
	\end{split}
\end{align}
where $\gamma_5=i\gamma^0\gamma^1\gamma^2\gamma^3$, and $\Sigma^j=\gamma^5\alpha^j$ are the $4\times 4$ spin matrices. 
With the use of Eq.~\eqref{0.4} expansion of the DHW-function can be written in the form \cite{IBB1}
\begin{align}\label{0.5}
	\begin{split}
 W(\bm{x},\bm{p},t)=&\frac{1}{4}(f_0+\gamma_5f_1-i\gamma^0\gamma_5f_2+\gamma^0f_3+
	\bm{\Sigma}\cdot\bm{g}_0+\bm{\alpha}\cdot\bm{g}_1\\
&	-i\bm{\gamma}\cdot\bm{g}_2+\gamma^0\bm{\Sigma}\cdot\bm{g}_3).
	\end{split}
\end{align}
The expansion coefficients are the same as $f_0,\,f_1,\,f_2,\,f_3$ and $\bm{g}_0,\,\bm{g}_1,\,\bm{g}_2,\,\bm{g}_3$ used in \cite{IBB1}.

Equations fulfilled by the expansion coefficients can be found by calculating their time derivatives using Dirac equation for the fermion field operators. In deriving these equations one usually adopts the Hartree-, or mean-EB approximation neglecting quantum fluctuations of the electromagnetic field \cite{IBB1,HAG}. This is equivalent to the replacements 
\begin{align*}
	\langle 0| \hat{F}^{\mu\nu}(\bm{x},t)\mathcal{U}(\bm{s},\bm{x},t)[\Psi(\bm{x}_1,t),\Psi^\dag(\bm{x}_2,t)]|0\rangle\rightarrow
	F^{\mu\nu}(\bm{x},t)	\langle 0| \mathcal{U}(\bm{s},\bm{x},t)[\Psi(\bm{x}_1,t),\Psi^\dag(\bm{x}_2,t)]|0\rangle,
\end{align*}
i.e., the operator of quantum electromagnetic field is replaced by classical C-number field. Application of the Dirac equation for fermion field 
operators in general case of space- and time-dependent electromagnetic field results in a complicated system of 16 integro-differential equations 
for the expansion coefficients of the DHW-function.  These equations simplify significantly for spatially homogeneous electric field vanishing for 
$t\rightarrow\pm\infty$, which is the subject of main interest in the present  paper. Initial conditions are determined by free vacuum value of 
DHW-function. It follows then from Eq.~\eqref{0.1} with zero electromagnetic field and free Dirac field operators that only the coefficients $f_3$ 
and $\bm{g}_1$ survive and their vacuum values are
\begin{align}\label{0.6}
	f^{vac}_3=-\frac{2mc^2}{E_{\bm p}},\quad \bm{g}^{vac}_1=-\frac{2c\bm{p}}{E_{\bm p}},
\end{align} 
where $E_{\bm p}=\sqrt{c^2\bm{p}^2+m^2c^4}$ is the free particle energy. In the case of spatially homogeneous electric field only the coefficients $\bm{g}_0$ 
and $\bm{g}_2$ couple to the vacuum values~\eqref{0.6}, so that it is sufficient to consider 10 equations for $f_3,\,\bm{g}_0,\,\bm{g}_1,\,\bm{g}_2$.  
They have the form~\cite{IBB2,archiwum}
\begin{align}\label{0.7}
	[\partial_t+e\bm{\mathcal{E}}(t)\cdot\nabla_{\bm p}]W(\bm{p},t)=\frac{c}{\hbar}M(\bm{p})W(\bm{p},t),
\end{align}
where $W$ denotes 10-dimensional vector
\begin{align}\label{0.7a}
W=[f_3,\,\bm{g}_0,\,\bm{g}_1,\,\bm{g}_2],
\end{align}
  and the $10\times 10$ matrix $M$ has the following block structure
\begin{align}\label{0.8}
	M(\bm{p})=\left[\begin{array}{cccc}
		0 & \bm{0}^T & \bm{0}^T & 2\bm{p}^T\\
		\bm{0} & \mathds{O}_3 & 2\bm{p}\times & \mathds{O}_3\\
		\bm{0} & 2\bm{p}\times & \mathds{O}_3 & -2mcI_3\\
		-2\bm{p} & \mathds{O}_3 & 2mcI_3 & \mathds{O}_3
	\end{array}\right],
\end{align}
where $\bm{0}$ and $\bm{p}$ are 3-dimensional null and momentum column vectors, $\mathds{O}_3$ -  the $3\times 3$ null matrix and $I_3$ is the 3-dimensional identity matrix. Notation $\bm{p}\times$ means that when acting on a 3-dimensional vector to the right it gives its vector product with $\bm{p}$. Explicitly,
\begin{align}\label{0.9}
	\bm{p}\times = \left[\begin{array}{ccc}
		0 & -p_3 & p_2\\
		p_3 & 0 & -p_1\\
		-p_2 & p_1 & 0
	\end{array}\right].
\end{align}

In closing this Section, we note that physical interpretation of the DHW-functions can be found in Ref.~\cite{IBB1}. In particular, the phase space energy density is given by \cite{IBB1, Blinne},
\begin{align}\label{0.9a}
\varepsilon(t, \bm{r}, \bm{p})=c\bm{p}\cdot\bm{g}_1(t, \bm{r}, \bm{p})+mc^2f_3(t, \bm{r}, \bm{p}).
\end{align}
The one particle distribution function, which will be used in Sect.~V for numerical analysis of momentum distributions, is defined as~\cite{Blinne}
\begin{align}\label{0.9b}
f(t,\bm{r}, \bm{p})=\frac{\varepsilon(t, \bm{r}, \bm{p})-\varepsilon_{vac}}{2E_p}=\frac{\varepsilon(t, \bm{r}, \bm{p})}{2E_p}+1,
\end{align}
where $\varepsilon_{vac}$ was expressed by vacuum DHW-functions~\eqref{0.6}. It is also worth noting that the DHW formalism
is very general as it allows one to account for an arbitrary electromagnetic field. However, for a spatially homogeneous electric field,
other approaches can be conveniently applied; one of which being developed next.

\section{Bispinorial representation of the DHW functions for spatially homogeneous electric field}
\label{bispinorial}

We consider Dirac equation in the spatially homogeneous electric field $\bm{\mathcal{E}}(t)=-\partial _t\bm{\mathcal{A}}(t)$, with the vector potential vanishing both for $t\rightarrow -\infty$ and $t\rightarrow \infty$.  Due to translational invariance of the problem spatial dependence of the wave function is of plane wave type,
\begin{align}\label{1.1}
	\Psi(t,\bm{x})=\exp\left(\frac{i}{\hbar}\bm{p}\cdot\bm{x}\right)\Phi_{\bm{p}r}(t),
\end{align}
where time-dependent bispinor $\Phi_{\bm{p}r}(t)$ is labeled by asymptotic momentum $\bm{p}$ and spin index $r$, and fulfills the equation
\begin{align}\label{1.2}
	i\hbar\partial_t\Phi_{\bm{p}r}(t)=H_D(t)\Phi_{\bm{p}r}(t),
\end{align}
where the time-dependent Hamiltonian reads
\begin{align}\label{1.3}
	H_D(t)=c\bm{\alpha}\cdot[\bm{p}-e\bm{\mathcal{A}}(t)]+\gamma^0mc^2.
\end{align}

To make contact with the DHW-functions we construct 16 expressions bilinear in the bispinor $\Phi_{\bm{p}r}(t)$
\begin{align}\label{1.4}
	S_a(\bm{p},t)=\sum_r\Phi^\dag_{\bm{p}r}(t)\Gamma_a\Phi_{\bm{p}r}(t).
\end{align}
Using the Dirac equation \eqref{1.2} and its Hermitian conjugate one finds equations fulfilled by the functions $S_a$,
\begin{align}\label{1.7}
	\partial_tS_a=\frac{i}{\hbar}\sum_r \Phi^\dag_{\bm{p}r}[H_D(t),\Gamma_a]\Phi_{\bm{p}r}.
\end{align}
The Dirac Hamiltonian $H_D(t)$ can be written in terms of $\Gamma$-matrices as
\begin{align}\label{1.8}
	H_D(t)=c\Gamma_{j+6}[p^j-e\mathcal{A}^j(t)]+mc^2\Gamma_3,
\end{align}
where summation convention for the Cartesian index $j$ was used. The $\Gamma$ matrices fulfill commutation relations
\begin{align}\label{1.9}
	[\Gamma_a,\,\Gamma_b]=i\sum_{c=0}^{15}f^c\,_{ab}\Gamma_c,
\end{align}
where $f^c\,_{ab}$ are real structure constants of the algebra of $\Gamma$ matrices. Substituting Eq.~\eqref{1.8} into Eq.~\eqref{1.7} and using Eq.~\eqref{1.9} gives
\begin{align}\label{1.10}
	\partial_tS_a=-\frac{c}{\hbar}(p^j-e\mathcal{A}^j)\sum_{b=0}^{15}f^b\,_{j+6,a}S_b-\frac{mc^2}{\hbar}\sum_{b=0}^{15}f^b\,_{3a}S_b.
\end{align}
Nonvanishing structure constants are (indices $i,j,k$ take the values $1,2,3$)
\begin{align*}\label{1.11}
	& f^k\,_{ij}=2\epsilon_{ijk},\\
	\nonumber  & f^{i+12} \, _{1,\,i+9}=2,\, f^{i+9} \, _{1,\,i+12}=-2,\,f^{i+12} \, _{2,\,i+6}=-2,\, f^{i+6} \, _{2,\,i+12}=2,\\
	& f^{i+9} \, _{3,\,i+6}=2,\, f^{i+6} \, _{3,\,i+9}=-2,\\
	& f^{k+3}\, _{i+3,j+3}=f^{k+3}\, _{i+6,j+6}=f^{k+3}\, _{i+9,j+9}=f^{k+3}\, _{i+12,j+12}=2\epsilon_{ijk},\numberthis \\
	& f^{k+6}\, _{i+3,j+6}=f^{k+9}\, _{i+3,j+9}=f^{k+12}\, _{i+3,j+12}=2\epsilon_{ijk},\\
	&f^1\,_{i+9,j+12}=2\delta_{ij},\,
	f^2\,_{i+6,j+12}=-2\delta_{ij},\,
	f^3\,_{i+6,j+9}=2\delta_{ij},
\end{align*}
plus the structure constants obtained from  anti-symmetry relation $f^a\, _{c\,b}=-f^a\, _{b\,c}$. It is now straightforward, though a little tedious, to derive 16 equations fulfilled by the functions $S_a$
\begin{subequations}\label{1.12}
	\begin{align}
	&	\partial_tS_0=0,\\
	& \partial_tS_1=-2\frac{mc^2}{\hbar}S_2,\\
	& \partial_tS_2=-2\frac{c}{\hbar}(p^j-e\mathcal{A}^j)S_{j+12}+2\frac{mc^2}{\hbar}S_1,\\
	& \partial_tS_3=2\frac{c}{\hbar}(p^j-e\mathcal{A}^j)S_{j+9},\\
	& \partial_tS_{k+3}=2\frac{c}{\hbar}\epsilon_{kjl}(p^j-e\mathcal{A}^j)S_{l+6},\\
	& \partial_tS_{k+6}=2\frac{c}{\hbar}\epsilon_{kjl}(p^j-e\mathcal{A}^j)S_{l+3}-2\frac{mc^2}{\hbar}S_{k+9},\\
	& \partial_tS_{k+9}=-2\frac{c}{\hbar}(p^k-e\mathcal{A}^k)S_3+2\frac{mc^2}{\hbar}S_{k+6},\\
	& \partial_tS_{k+12}=2\frac{c}{\hbar}(p^k-e\mathcal{A}^k)S_2.
	\end{align}
\end{subequations}
Note that equations containing 6 functions $S_0,\,S_1,\,S_2,\,S_{13},\,S_{14},\,S_{15}$ do not couple to remaining ten equations for $S_3,\,S_4,\,S_5,\,S_6,\,S_7,\,S_8,\,S_9,\,S_{10},\,S_{11},\,S_{12}$. Denoting
\begin{align}\label{1.13}
S_3=h_3,\quad (S_4,S_5,S_6)=\bm{h}_0,\quad  
	& (S_7,S_8,S_9)=\bm{h}_1, \quad
	(S_{10},S_{11},S_{12})=\bm{h}_2,
\end{align}
we see that equations \eqref{1.12} for  ten-dimensional vector $V=[h_3,\,\bm{h}_0,\,\bm{h}_1,\,\bm{h}_2]$ can be written in the matrix form as
\begin{align}\label{1.14}
	\partial_tV=\frac{c}{\hbar}M(\bm{p}(t))V,
\end{align}
where
\begin{align}\label{1.15}
	\bm{p}(t)=\bm{p}-e\bm{\mathcal{A}}(t),
\end{align}
and the matrix $M$ is given by~\eqref{0.8}. The same system of ordinary differential equations follows from Eq.~\eqref{0.7} after applying the method 
of characteristics to first order partial differential equations~\cite{IBB1,HAG,archiwum,Blinne}. Therefore two vectors $W$ and $V$ obey the same system of 
ordinary differential equations. In order to identify fully $V$ and $W$ one needs to show that they fulfill also the same initial conditions, which for $W$ 
are given by Eq.~\eqref{0.6} and $\bm{g}_0^{vac}=0=\bm{g}_2^{vac}$. 

The Dirac wave function pertaining to pair creation process should fulfill Feynman boundary conditions: for $t\rightarrow -\infty$ it contains only solutions of free Dirac equation with negative energy, whereas for $t\rightarrow\infty$ it is a combination of positive and negative energy parts with the negative energy contribution equal to the wave function of the created positron. Extensive discussion of boundary conditions fulfilled by solutions of Dirac equation in classical electromagnetic field can be found in \cite{IBB3}.
It can be also shown that Feynman boundary conditions are ``forced'' by LSZ-reduction formulae for the S-matrix element of pair creation.
For $t\rightarrow -\infty$ we have therefore
\begin{align}\label{1.16}
	\Phi_{\bm{p}s}=\exp\left(\frac{i}{\hbar}E_{\bm p}t\right)w^{(-)}_{-\bm{p}s}.
\end{align}
Substituting Eq.~\eqref{1.16} into Eq.~\eqref{1.4} one can show that coefficients \eqref{1.13} fulfill the following initial conditions for $t\rightarrow -\infty$
\begin{align}
	\bm{h}_0^{0}=0=\bm{h}_2^{0},\quad	h^{0}_3=-\frac{2mc^2}{E_{\bm p}},\quad \bm{h}^{0}_1=-\frac{2c\bm{p}}{E_{\bm p}},
\end{align} 
corresponding exactly to vacuum initial conditions for the vector $W$.

The bispinorial approach to the dynamic Sauter-Schwinger pair production by spatially homogeneous 
electric fields has been developed in this Section. Importantly, it has been proven to be equivalent to the DHW formalism
described in Sect.~II. Therefore, similar to the DHW method, it has advantaged over other approaches. Specifically, it allows 
to treat an arbitrarily polarized time-dependent electric field. Having said that, we turn to the case of a linear polarization, 
for which other well-established theories exist and can be tested against (see, for instance, Refs.~\cite{KMK,KK2019} and references therein).

\section{Linearly polarized field and analogy with two level atom}
\label{linear}
In general, the vector $W$ (or, equivalently $V$) can be expressed as a combination of ten orthonormal basis vectors $\mathds{E}_a$
\begin{align}\label{3.1}
	W=-2\sum_{a=1}^{10}u_a\mathds{E}_a.
\end{align}
With choice of $-\frac{1}{2}W^{vac}$ as one of the basis elements one can show that for linearly polarized field $\bm{\mathcal{A}}(t)=\mathcal{A}(t)\bm{n}$ three vectors
\begin{align}\label{3.2}
 \mathds{E}_1=\frac{c}{E_{\bm p}\epsilon_\perp}
 \left[\begin{array}{c}
 	 -mc^2(\bm{n}\cdot\bm{p})\\ \bm{0}\\ \dfrac{E^2_{\bm p}\bm{n}}{c}-c(\bm{n}\cdot\bm{p})\bm{p}\\ \bm{0}
 \end{array}\right],\,\mathds{E}_2=\frac{c}{\epsilon_\perp} \left[\begin{array}{c}
 0\\ \bm{p}\times\bm{n}\\ \bm{0}\\ mc\bm{n}
\end{array}\right],\,\mathds{E}_3=\frac{1}{E_{\bm p}}\left[\begin{array}{c}
mc^2\\ \bm{0}\\ c\bm{p}\\ \bm{0}
\end{array}\right],
\end{align}
where $\epsilon_\perp=\sqrt{c^2\bm{p}_\perp^2+m^2c^4}$, form a set closed under the action of $\bm{n}\cdot\nabla_{\bm p}$ and $M$ in Eq.~\eqref{0.7}. 
Note that $\mathds{E}_3=-(1/2)W^{vac}$. Choosing $\bm{n}$ in the $z$-direction (${\bm n}={\bm e}_3$) we have
\begin{align}\label{3.3}
	\begin{split}
		&\frac{\partial}{\partial p_3}\mathds{E}_1=-\frac{c\epsilon_\perp}{E_{\bm p}^2}\mathds{E}_3,\quad\frac{\partial}{\partial p_3}\mathds{E}_2=0,\quad\frac{\partial}{\partial p_3}\mathds{E}_3=\frac{c\epsilon_\perp}{E_{\bm p}^2}\mathds{E}_1,\quad \\
		&M\mathds{E}_1=\frac{2E_{\bm p}}{c}\mathds{E}_3,\quad M\mathds{E}_2=-\frac{2E_{\bm p}}{c}\mathds{E}_1,\quad M\mathds{E}_3=0.
	\end{split}
\end{align}
Solution of Eq.~\eqref{0.7} can be expressed as
\begin{align}\label{3.4}
		W(\bm{p},t)=-2\sum_{a=1}^3u_a(\bm{p},t)\mathds{E}_a(\bm{p}(t)).
\end{align}
Substituting Eq.~\eqref{3.4} to Eq.~\eqref{1.14}, and denoting $[u_1,\,u_2,\,u_3]=\bm{u}$ we obtain the precession-type equation for $\bm{u}$
\begin{align}\label{3.5}
	\partial_t\bm{u}=\bm{a}\times\bm{u},
\end{align}
with the vector $\bm{a}$ given by
\begin{align}\label{3.6}
	\bm{a}=[0,\,-2\Omega_{\bm p}(t),\,2\omega_{\bm p}(t)],
\end{align}
where
\begin{align}\label{3.7}
	\omega_{\bm p}(t)=\frac{E_{{\bm p}(t)}}{\hbar}=\frac{1}{\hbar}\sqrt{c^2\bm{p}_\perp^2+c^2[p_3^2-e\mathcal{A}(t)]^2+m^2c^4},\quad \Omega_{\bm p}(t)=\frac{ce\epsilon_\perp\mathcal{E}(t)}{2E_{{\bm p}(t)}^2},
\end{align}
and where the temporal dependence of the electric field is given by ${\cal E}(t)=-\dot{\cal A}(t)$. Note that the initial condition for $\bm{u}$ has the form $\bm{u}^{vac}=[0,\,0,\,1]$.	

Three equations \eqref{3.5} can be reduced to a system of two equations by expressing $\bm{u}$ in the form of spinorial decomposition, analogous to that used in~\cite{IBB2,archiwum}
\begin{align}\label{3.8}
	\bm{u}=\chi^\dag\bm{\sigma}\chi,
\end{align}
where $\chi$ is a two-component spinor and $\bm{\sigma}$ are Pauli matrices. Substitution of Eq.~\eqref{3.8} to~\eqref{3.5} leads to the equation for $\chi$ which has the same structure 
as the Schr\"{o}dinger equation describing time evolution of a two-level atom. This equation has been derived in the context of pair-creation by different method previously (see, e.g., Refs.~\cite{IBB2,KMK} and references therein),
\renewcommand{\arraystretch}{1.5}
\begin{align}\label{3.10}
	i\partial_t\left[\begin{array}{c}
		c_{\bm{p}}^{(1)}(t)\\c_{\bm{p}}^{(2)}(t)\end{array}\right]=\left[\begin{array}{cc}
			\omega_{\bm{p}}(t) & i\Omega_{\bm{p}}(t)\\-i\Omega_{\bm{p}}(t) & -\omega_{\bm{p}}(t)
	\end{array}\right]\left[\begin{array}{c}
	c_{\bm{p}}^{(1)}(t)\\c_{\bm{p}}^{(2)}(t)\end{array}\right],
\end{align}
where $c_{\bm{p}}^{(1)}(t)$ and $c_{\bm{p}}^{(2)}(t)$ are, respectively, upper and lower components of $\chi$. Initial conditions read
\begin{align}\label{3.11}
	c_{\bm{p}}^{(1)}\vert_{t\rightarrow -\infty}=1,\quad	c_{\bm{p}}^{(2)}\vert_{t\rightarrow -\infty}=0.
\end{align}
Third component of $\bm{u}$ is equal to $|c^{(1)}_{\bm{p}}|^2-|c^{(2)}_{\bm{p}}|^2$ and for two level atom corresponds to "population inversion" 
(with opposite sign). Before action of the electric field $u_3=1$, which corresponds to the vacuum state with no pairs. During action of the electric 
field $e^+e^-$ pairs are created so that $|c^{(1)}_{\bm{p}}|^2<1$ and $|c^{(2)}_{\bm{p}}|^2>0$ with $|c^{(1)}_{\bm{p}}(t)|^2+|c^{(2)}_{\bm{p}}(t)|^2=1$. 
Hence, $|c^{(2)}_{\bm{p}}|^2$ for $t\rightarrow\infty$ can be interpreted as  momentum distribution of created fermionic pairs $f(\bm{p})$ \eqref{0.9b}.
Explicitly,
\begin{align}\label{0.9c}
 	f(\bm{p})=1-u_3=2|c^{(2)}_{\bm{p}}|^2.
\end{align}
Let us note in closing this section that a similar to \eqref{3.10} system of equations  can be derived for bosons by applying other, then those 
based on the Wigner formalism, methods of QED (see, e.g., Ref.~\cite{KK2019} and references therein). However, in this case the time-evolution is pseudounitary.

From now on we use units in which $\hbar=1$. Moreover, $m$ and $e$ will refer to the electron rest mass and charge, respectively.

\section{Threshold effects and \kolorb{vortices}}
\label{sec:threshold}

\begin{figure}
  \includegraphics[width=0.9\linewidth]{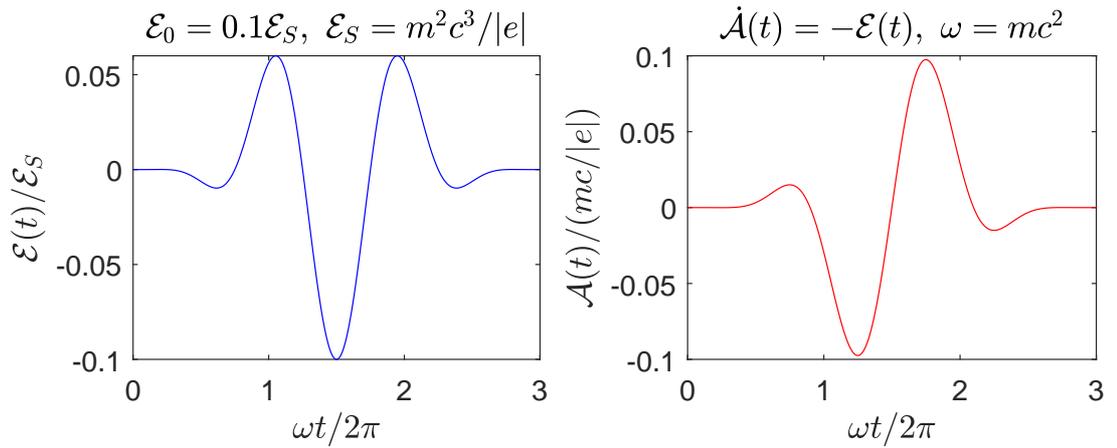}
\caption{Time-dependent electric field strength $\mathcal{E}(t)$ for $\mathcal{E}_0=0.1\mathcal{E}_S$ and $N=3$, as defined by Eq.~\eqref{th1}, and the corresponding vector potential function $\mathcal{A}(t)$. Contrary to the electric field, the amplitude of the vector potential depends on the frequency $\omega$. 
}
\label{fepole}
\end{figure}

In our further investigations we choose the electric field $\mathcal{E}(t)$ such that
\begin{equation}
\mathcal{E}(t)=\begin{cases}\mathcal{E}_0\sin^4\bigl( \frac{1}{2N}\phi\bigr)\cos(\phi), & \phi\in [0,2\pi N], \cr
0, & \phi\notin [0,2\pi N], \end{cases}
\label{th1}
\end{equation}
where $\phi=\omega t$ and $N=3$. The integer $N$ determines the number of cycles within the electric field pulse and for $N\geqslant 3$ the condition
\begin{equation}
\int_{-\infty}^{\infty}\mathrm{d}t \,\mathcal{E}(t)=0
\label{th2}
\end{equation}
is satisfied. Due to this property the vector potential function,
\begin{equation}
\mathcal{A}(t)=-\int_{-\infty}^t \mathrm{d}\tau \,\mathcal{E}(\tau),
\label{th3}
\end{equation}
can be chosen such that it vanishes both in the remote past and in the far future,
\begin{equation}
\lim_{t\rightarrow\pm\infty}\mathcal{A}(t)=0.
\label{th4}
\end{equation}
The shapes of both functions for $\mathcal{E}_0=0.1\mathcal{E}_S$ and $\omega=mc^2$ are presented in Fig.~\ref{fepole}, where $\mathcal{E}_S=m^2c^3/|e|$ 
is the Sauter-Schwinger electric field strength~\cite{Sauter,Sauter2,Schwinger}. For the electron momentum vector, we will separate its parallel and 
perpendicular components as measured with respect to the direction of the electric field oscillations ${\bm e}_3$ such that $\bm{p}=p_{\bot}\bm{e}_{\bot}+p_{\|}\bm{e}_3$, 
where $\bm{e}_{\bot}$ is a unit vector perpendicular to $\bm{e}_3$.

\begin{figure}
  \includegraphics[width=0.9\linewidth]{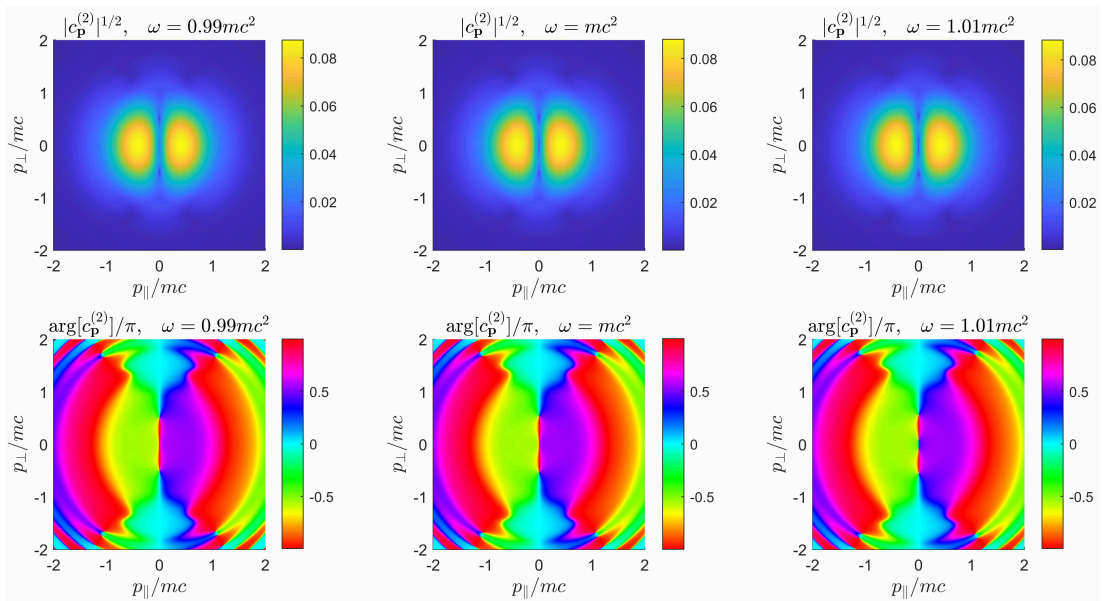}
\caption{Momentum distributions of electrons created from the QED vacuum by the electric field, illustrated in Fig.~\ref{fepole}. In the upper row, 
the distributions $|c_{\bm{p}}^{(2)}|^{1/2}$ (the power $1/2$ is chosen for visual purposes) are presented for three chosen frequencies $\omega$
(equivalent to photon energies). In the lower row, the corresponding phases of $c_{\bm{p}}^{(2)}$ are demonstrated.
}
\label{dynpair1}
\end{figure}

As was mentioned in Introduction, the process of creating electron-positron pairs in QED has many analogies with ionization of atoms, in which the role 
of the time-dependent electric field is played by a strong laser pulse in the dipole approximation. For this case the concept of photons as quanta 
of energy absorbed or emitted by the system is commonly used. One can then talk about multiphoton ionization and an energy threshold for that process. 
Moreover, such a threshold is dynamically increased as the electric field becomes stronger, which leads to the so-called threshold effects and channel closing in ionization \cite{KFS2006}. It turns out that in the case of the dynamic Sauter-Schwinger process this heuristic picture can also be applied in order to describe qualitative changes in momentum distributions of the created particles (as, for instance, in the coherent energy combs studied in Refs.~\cite{KMK,Dunne}). This can be done even for the very short pulses considered here.

\begin{figure}
  \includegraphics[width=0.9\linewidth]{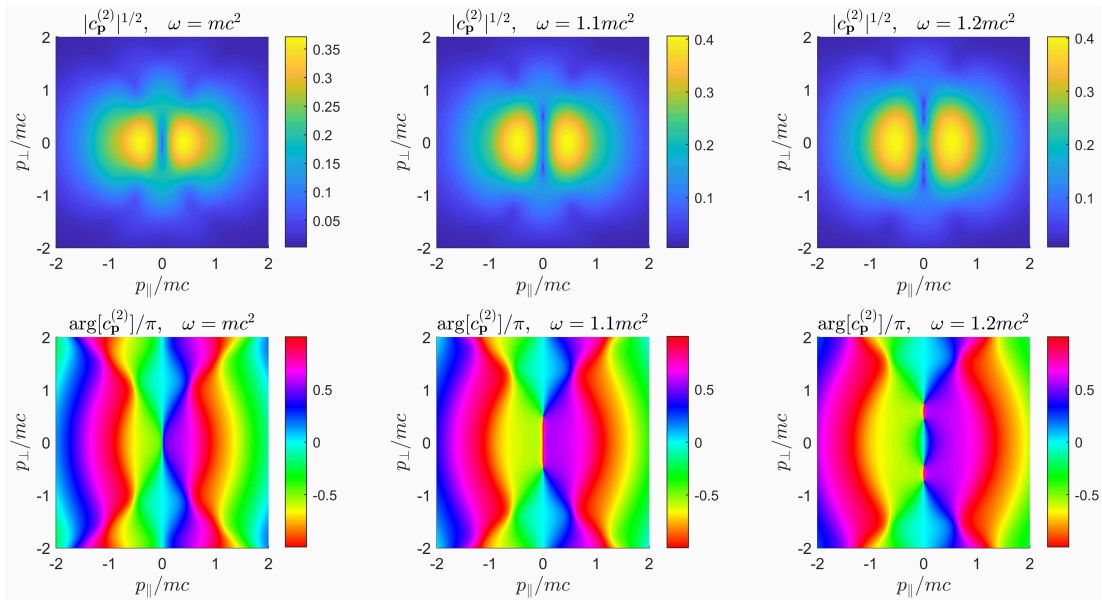}
\caption{The same as in Fig.~\ref{dynpair1}, but for lager electric field amplitude $\mathcal{E}_0=0.5\mathcal{E}_S$ and lager frequencies.
}
\label{dynpair2}
\end{figure}

Another interesting effect, which appears as a result of a time-dependent electric field interacting with QED vacuum, is the creation or annihilation 
of vortex lines in the electron momentum distributions. The properties of vortex lines and their entanglement were thoroughly analyzed in 
Refs.~\cite{BS,Taylor}. In both photoionization and photodetachment, the creation and annihilation of vortex lines were studied for linearly \cite{Pe1} 
and circularly \cite{Pe2,Pe3,Felipe,Majczak} polarized fields. It has been shown how the time-reversal symmetry of the laser pulse leads to the annihilation of vortex-antivortex pairs and the creation of spirals in momentum distributions \cite{Majczak}. Note that such spirals have been predicted theoretically in \cite{SAF} and confirmed experimentally in \cite{Wollen}. Moreover, the application of the DHW-function formalism allowed to show that similar spiral structures also appear in pair creation by a train of two circularly polarized electric field pulses of opposite helicity \cite{Xie1,Xie2}. Therefore, the question arises: Can vortices be expected in momentum distributions of created pairs for linearly polarized electric field pulses?

To address this question, in Fig.~\ref{dynpair1} we present the momentum distributions of electrons created by a linearly polarized electric field pulse of
different frequencies, which have been selected close to 
the `two-photon' threshold of pair creation. For $\omega=0.99mc^2$ (i.e., just before opening the `two-photon' channel) we observe two singular points 
for which the amplitude $c_{\bm{p}}^{(2)}$ vanishes and the phase ${\rm arg}[c_{\bm{p}}^{(2)}]$ cannot be uniquely defined. Because of the axial 
symmetry of the problem, it can be concluded that these two points belong to the same vortex line. In the current case, the latter is represented by 
a circle in the three-dimensional momentum space. In fact, one can even define the orientation of this closed line by exploiting the analogies with 
the circuit along which the electric current flows and generates, according to the Amper\'e's law, the vortex-type magnetic field. To this end, 
let us define the `magnetic field' $\bm{B}(\bm{p})$ such that
\begin{equation}
\bm{B}(\bm{p})=\bm{\nabla}_{\bm{p}}\bigl({\rm arg}[c_{\bm{p}}^{(2)}] \bigr).
\label{jke1}
\end{equation}
Its circulation around a singular point is $\pm 2\pi$, hence the `electric current' becomes $I=\pm 2\pi$, if we put the magnetic permeability $\mu_0=1$. 
In particular, for $\omega=0.99mc^2$ (left panel in Fig.~\ref{dynpair1}), we have `$-$' for $(p_{\|}=0,p_{\bot}>0)$ and the current flows behind the plane, 
whereas for $(p_{\|}=0,p_{\bot}<0)$ we have `$+$' and the current flows towards the reader. Thus the orientation of the vortex line can be uniquely 
attributed to the direction of the `electric current'. As the frequency $\omega$ increases, we observe the appearance of a new vortex line. 
The case of $\omega=mc^2$ (which is the threshold frequency for the two-photon pair creation) corresponds to a transition in which, for 
$\bm{p} = \bm{0}$, the radius of the new vortex circle is close to zero (middle column in Fig.~\ref{dynpair1}). After exceeding this value 
(the case of $\omega = 1.01mc^2$, i.e., the right column in Fig.~\ref{dynpair1}), the second circular vortex line appears, with an orientation opposite 
to the previous one (bottom panels). While increasing the frequency $\omega$, the radii of both circular vortex lines also grow. 
This, in turn, results in the merging of the two well-defined lobes of high probability into a single structure, the maximum of which is found at zero 
momentum. This situation is discussed in Sect.~\ref{sec:lt}.

In Fig.~\ref{dynpair2} we demonstrate the same phenomenon, but for a larger amplitude of the electric field. The only significant difference is that 
now the threshold frequency for the two-photon pair creation is shifted upwards and its value is between $1.1mc^2$ (one vortex line) and $1.2mc^2$ 
(two vortex lines). A plausible interpretation of this fact can be based on analogies with the multiphoton ionization in which, for larger 
intensity of the electromagnetic field, the so-called ponderomotive shift of the threshold energy is observed~\cite{KFS2006}. Similar effects, 
but in the context of photodetachment by circularly polarized laser pulses, have been discussed in Ref.~\cite{Felipe}.

\begin{figure}
  \includegraphics[width=0.9\linewidth]{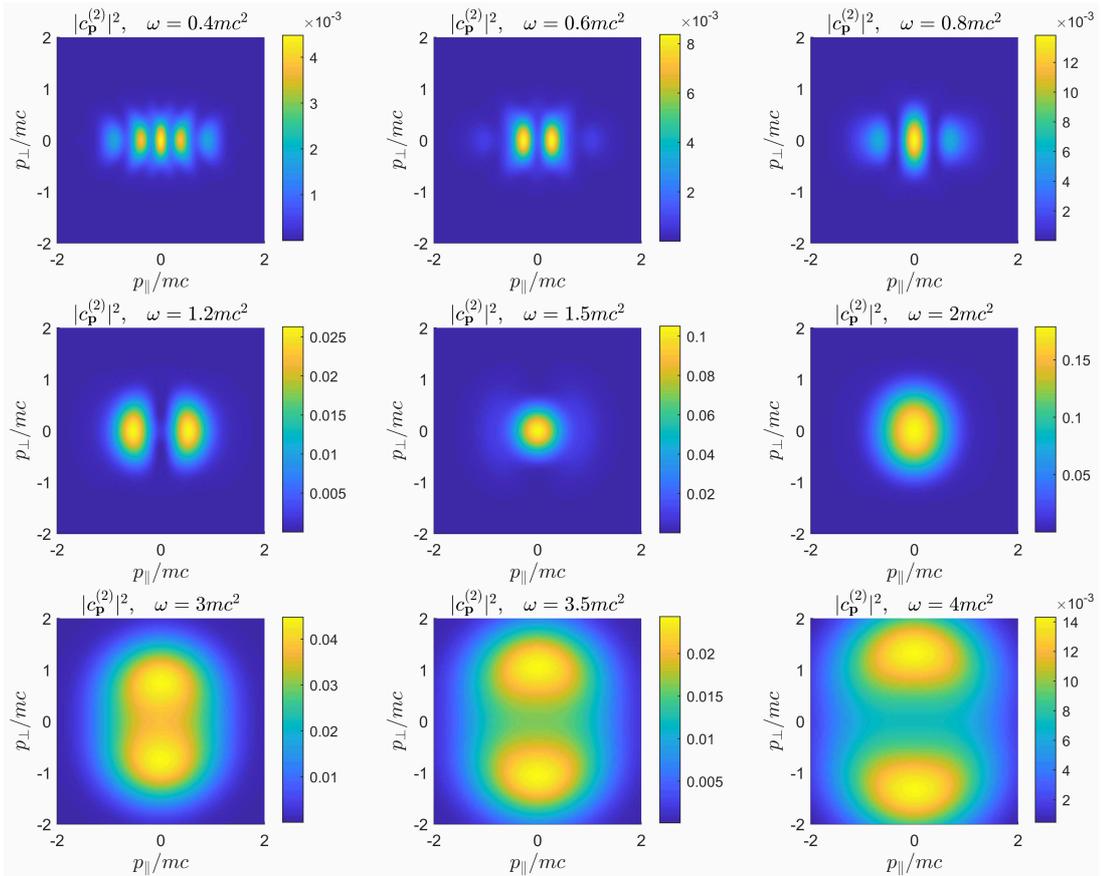}
\caption{Momentum distributions of electrons created by oscillating electric field for different frequencies and for the electric field amplitude $\mathcal{E}_0=0.5\mathcal{E}_S$. Starting from the one-photon threshold frequency, $\omega=2mc^2$, a qualitative change in the shapes of the high-probability structures is observed.
}
\label{dynpairmom2}
\end{figure}

\section{Longitudinal and transverse momentum sharing}
\label{sec:lt}

The Schwinger formula for probability rate of pair production per unit volume in the case of a constant (or slowly-changing-in-time) electric field can be derived using the tunneling 
formalism~\cite{LR}. According to this formula, an increment of the perpendicular momentum of the particles, $|{\bm p}_\perp|$, is accompanied 
by a rapidly vanishing creation rate. However, for rapidly changing fields the tunneling theory is no longer applicable. This is supported by the 
analysis presented above, as the momentum distributions for pair production are not elongated in the direction of the electric field. As it has been 
shown in Ref.~\cite{KK012104}, for sufficiently high frequencies $\omega$ particles prefer to be created in the direction perpendicular to the electric 
field. This counterintuitive phenomenon is illustrated in Fig.~\ref{dynpairmom2}. For low frequencies, $\omega \leqslant 0.8mc^2$, the distributions 
are concentrated around the axis of vanishing transverse momentum. However, as the frequency increases, the distributions begin to concentrate around 
zero momentum. This happens until the one-photon threshold is reached. A further increase of frequency causes the position of the high-probability 
regions in the distribution to migrate towards the direction perpendicular to the electric field (i.e., towards larger $p_\perp$). Furthermore, at 
$\omega=4mc^2$ the high-probability zone, in the three-dimensional space, takes the shape of a torus centered around the 
${\bm p}_\perp={\bm 0}$. This means that under such conditions particles prefer to be ejected in the direction perpendicular to the electric field 
vector. As shown in Ref.~\cite{KK012104}, the distribution for pair creation, when integrated over particles momenta, starts to saturate 
(or even decreases) with the increasing frequency, leading to the seemingly unexpected stabilization phenomenon. In fact, the stabilization effects 
appear to be quite common in the strong-field QED, as discussed for instance in Refs.~\cite{KKE,LR2,ER21,Kemp1,Kemp2}.

In summary, although we have concentrated our discussion on the fermionic distribution function $f(\bf{p})$ and the phase of the momentum amplitude $c^{(2)}_{\bm{p}}$, the other components of the DHW functions \eqref{0.7a} can be also determined by applying Eqs.~\eqref{3.8}, \eqref{3.4} and \eqref{3.2}. This topic is, however, beyond the scope of the present work and is going to be considered in due course.

\section{Conclusions}
\label{sec:conclusions}

In this paper, we have formulated the bispinorial approach to $e^-e^+$ pair production in spatially homogeneous electric
fields. The method has turned out to be equivalent to the DHW formalism, that was introduced in Ref.~\cite{IBB1}.
We have shown that the Sauter-Schwinger
pair production by a linearly polarized time-dependent electric field reduces formally to solving a two-level model in compliance with 
Ref.~\cite{KMK} (see also, references therein). The advantage of this approach is that one gains access to probability amplitudes
and, therefore, to their phases. The latter allow to uniquely identify vortices and antivortices in momentum distributions of
created pairs. As it has been demonstrated in our paper, for a linearly polarized pulsed electric field they appear in pairs.
We have also analyzed the vortex patterns while increasing the field frequency across the two-photon threshold. While we have observed new vortex-antivortex pair, the general features of momentum distributions also change across the threshold. Specifically, we have seen that below the one-photon threshold the particles are created most efficiently along the polarization direction of the electric
field, whereas above the threshold -- in the perpendicular direction. This shows different characteristics of the Sauter-Schwinger
process while passing from low- to high-frequency regimes of electric-field--vacuum interactions.

\section*{Acknowledgements}
The work of (F.C.V., K.K., and J.Z.K.) was supported by the National Science Centre (Poland) under Grant 
No. 2018/31/B/ST2/01251.

\end{document}